\documentstyle[12pt,bezier]{article}

\newcommand{\newsection}{
\setcounter{equation}{0}
\section}

\def\appendix#1{
  \addtocounter{section}{1}
  \setcounter{equation}{0}
  \renewcommand{\thesection}{\Alph{section}}
  \section*{Appendix \thesection\protect\indent \parbox[t]{11.715cm} {#1}}
  \addcontentsline{toc}{section}{Appendix \thesection\ \ \ #1}
  }

\newcommand{\tr}[1]{\,{\rm Tr}\,#1}

\def\e{{\,\rm e}\,}
\def\impart{{\rm Im}}
\def\tmax{\theta _{\rm max}}
\def\realpart{{\rm Re}}
\def\diag{{\rm diag}}
\def\meff2{m^2_{{\rm eff}}}
\newcommand{\rf}[1]{(\ref{#1})}

\begin{document}

\begin{titlepage}
\begin{flushright}
SMI-96-4 \\
hep-th/9606117\\
\today
\end{flushright}
\vspace{.5cm}

\begin{center}
{\LARGE Adjoint non-Abelian Coulomb gas at large N}
\end{center} \vspace{1cm}
\begin{center} {\large G.W.\ Semenoff}
\footnote{E--mail:  \ semenoff@physics.ubc.ca \ /
\ semenoff@nbivms.nbi.dk \ }
\\ \mbox{} \\
{\it Department of Physics, University of British Columbia,} \\ {\it
6224 Agricultural Road, }\\{\it Vancouver, British Columbia, Canada
V6T 1Z1 }
\\
\vspace{0.5cm} \mbox{} \\ {\large and} \\
\vspace{0.5cm} \mbox{} \\
{\large K.\ Zarembo}
\footnote{E--mail:   \ zarembo@class.mian.su \ }\\
 \mbox{} \\ {\it Steklov Mathematical Institute,} \\
{\it Vavilov Street 42, GSP-1, 117966 Moscow, RF}
\\ \vskip .2 cm
and  \\  \vskip .2 cm
{\it Institute of Theoretical and Experimental Physics,}
\\ {\it B. Cheremushkinskaya 25, 117259 Moscow, RF} \\
\end{center}

\vskip 1 cm
\begin{abstract}
The non-Abelian analog of the classical Coulomb gas is discussed.  The
statistical mechanics of arrays of classical particles which transform
under various representations of a non-Abelian gauge group and which
interact through non-Abelian electric fields are considered.  The
problem is formulated on the lattice and, for the case of adjoint
charges, it is solved in the large N limit.  The explicit solution
exhibits a first order confinement-de-confinement phase transition
with computable properties.  In one dimension, the solution has a
continuum limit which describes 1+1-dimensional quantum chromodynamics
(QCD) with heavy adjoint matter.
\end{abstract}

\vspace{1cm}
\noindent

\end{titlepage}
\setcounter{page}{2}

\section{Introduction}

The classical Coulomb gas is an important model in statistical
mechanics.  It is solvable in one dimension and in two dimensions it
exhibits interesting critical phenomena.  In this Paper, we shall
formulate a non-Abelian generalization of the Coulomb gas.  We
consider the thermodynamic properties of arrays of non-dynamical
quarks which transform under irreducible representations of the gauge
group and which interact via non-Abelian electric fields.  The main
motivation is to study the confinement-deconfinement phase transition
which is expected to occur in gauge theory at sufficiently high
temperature or density.

The model which we formulate is most interesting in the case where the
quarks transform under the adjoint representation of the gauge group.
In that case, it can be solved explicitly in the large N limit and
exhibits a first order phase transition when the spatial dimension is
D=1,2,3.  When D=1, it has a continuum limit where it represents
(1+1)--dimensional QCD with very heavy adjoint matter fields.

In the absence of quarks, i.e. the zero density limit, the model which
we consider reduces to the strong coupling limit of lattice Yang-Mills
theory.  In that limit, only the color-electric interactions are
retained.  The phase structure of that system was studied in the large
$N$ limit in refs.\cite{Zar,Boul}.  There, it was found that in D=2
and D=3, there is a first order phase transition which occurs as the
parameter $\gamma=e^2N/2T$ is varied.  The strong coupling, low
temperature, confining phase occurs at $\gamma>>1$ and the weak
coupling, high temperature, de-confined phase at $\gamma<<1$.  In the
former case one can find an explicit solution of the lattice theory in
the large $N$ limit.  This phase is stable to small fluctuations for
$\gamma$ greater than some critical value. In the small $\gamma$
limit, approximate techniques are required and a de-confined phase is
found.  There is a co-existence region where both phases are stable to
small fluctuations, and are therefore separated by an energy barrier.
In that region, there is a first order phase transition between them.

In $D=2,3$, the strong coupling theory is not renormalizable.  The
phase transition is first order, and there is no limit where the
latent heat is small, i.e. of order $N^2$ times a quantity which
remains finite as the lattice spacing is taken to zero.  One could
speculate that, if the strong coupling limit were corrected to include
the magnetic term in the Hamiltonian, one would recover a
renormalizable theory, manifest in the fact that there would be some
limit in which the first order phase transition had finite latent heat
in the continuum.  There are good arguments for this scenario in the
literature.  As well, large $N$ techniques have been used to study the
confinement-deconfinement transition in pure Yang-Mills
theory~\cite{BCDMP}.

In D=1, there is no magnetic term in the full Hamiltonian, and
Yang-Mills theory can be solved exactly using a technique similar to
the strong coupling approximation.  Rather than being generated
dynamically, the confining quark-antiquark potential appears at the
tree level where the string tension is proportional to the coupling
constant $e^2$.  However, in D=1, pure Yang-Mills theory is trivial in
that it has no propagating degrees of freedom.  It therefore does not
exhibit a phase transition when the temperature is non-zero and it is
in the confining phase for all values of $\gamma$.\footnote{ 1+1
dimensional Yang-Mills theory on the sphere is known to have a third
order Kazakov-Douglas\cite{dk} phase transition.  However, in the
cylindrical geometry which is appropriate to finite temperature
Yang-Mills theory on the open line, $R^1$, there is no third order
phase transition.}

In this paper, we consider the model where a gas of charged sources
(quarks) in the adjoint representation of the gauge group are added to
strongly coupled Yang-Mills theory.  For $D=2,3$ there is still a
first order phase transition which occurs on a critical line in the
$\gamma-\lambda$ plane, where $\lambda \sim e^{-\mu/T}$ is the
fugacity and $\mu$ is the chemical potential of the quarks.  When
$\gamma$ is greater than a certain critical value, the deconfinement
transition can be induced by increasing the density, which is
controlled by increasing $\lambda$. There is also a third order phase
transition in the de-confined phase, similar to the Gross-Witten
~\cite{GW} transition of D=1 lattice Yang-Mills theory.

In D=1, the coupling of sources to Yang-Mills theory makes the model
non-trivial.  The continuum version of this model has been studied in
ref.\cite{stz}.  There is a first order phase transition between the
confining and de-confined phases which can be obtained by increasing
$\lambda$ for any value of $\gamma$.  This phase transition originates
in a percolation transition for electric flux lines.  The quantum
states are arrays of adjoint quarks on the line with non-dynamical
strings of electric flux joining them.  Each quark must have one line
of flux entering and one line leaving it.  For a fixed number of
quarks, the model is explicitly solvable and has a finite dimensional
Hilbert space, corresponding with the different ways of distributing
electric flux so that the states are gauge invariant.  In the large
$N$ limit, the energy of a state is proportional to the total length
of lines of electric flux plus the chemical potential times the number
of quarks.  At low temperature and density, the statistical sum is
dominated by configurations which are a dilute gas of mesons -- color
neutral bound states of two or more adjoint quarks bound together by
the appropriate number of strings of electric flux.

The property of confinement is characterized by measuring the energy
necessary to insert a fundamental representation quark-antiquark pair
into the system.  Gauge invariance requires that the pair is connected
by a single string of electric flux.  At low temperature, the energy
of this string is proportional to its length, leading to the linear,
confining interaction.  In a deconfined phase, for large separation,
the energy would be a constant as the distance between the quark and
antiquark is varied.  This occurs when, at the deconfinement phase
transition, the strings percolate in the one--dimensional space, and
the addition of a flux string between a quark and an antiquark adds a
small amount (or no) energy to the energy of the typical
configuration.

In the next few subsections, we review some of the formalism which we
require in the later sections - the Hamiltonian formalism of lattice
Yang-Mills theory, the construction of the partition function in that
formalism and some properties of states in the non-Abelian Coulomb
gas.

\subsection{Hamiltonian formulation of
lattice Yang-Mills theory with external sources}

We begin by reviewing the Hamiltonian formalism of Yang-Mills theory
on the lattice. This is standard material which can be found in
ref.\cite{kogut}, for example.  In the Hamiltonian formalism, time is
continuous and the space is approximated by a hypercubic lattice with
sites $x,y,\ldots$ and oriented links $\l$.  The spatial gauge fields
are operator valued unitary matrices, $U_\l$, and electric fields are
operator valued Lie algebra elements, $E_\l$,
\begin{equation}
U_\l U^{\dagger}_\l~=~1~=~U^{\dagger}_\l U_\l
~~,~~
E_\l^{\dagger}=E_\l
\end{equation}
Both are associated with oriented links and obey the reflection
conditions
\begin{equation}
U_{-\l}=U_\l^{\dagger}~~,~~E_{-\l}=-U_{\l}^{\dagger}E_{\l}U_{\l}
\end{equation}
The electric fields can be expanded in a basis
\begin{equation}
E_\l = \sum_{A} E^A_\l T^A
\end{equation}
where $T^A$ are hermitean generators in the fundamental representation
of the Lie algebra
\begin{equation}
\left[ T^A , T^B\right]=if^{ABC}T^C
\end{equation}
The operators have the algebra
\begin{equation}
\left[ U_\l,U_{\l'}\right]=0~~,~~\left[ E^A_\l, E^B_{\l'}
\right]= if^{ABC}E^C_\l \delta_{\l \l'}
~~,~~
\left[ E_\l^A, U_{\l'} \right]=T^A U_\l \delta_{\l \l'}
\label{commutator}
\end{equation}
The Hamiltonian is
\begin{equation}
H=\sum_{\l,A}\frac{e^2}{2}\left( E^A_\l\right)^2 +
\sum_{\Box}\frac{1}{2e^2}{\rm Tr}\left( \prod_{\Box}U
+\prod_{\Box}U^{\dagger}\right)
\label{ham}
\end{equation}
and the gauge constraint is
\begin{equation}
{\cal G}^A(x)\equiv \sum_{\l\in n(x)}E_\l^A + \sum_{i=1}^K
~T^A_{R_i}~\delta(x-x_i)~\sim~0
\label{gl}
\end{equation}
The first term in the Hamiltonian is the electric energy and the
second term, which is summed over oriented elementary plaquettes
$\Box$, is the magnetic energy.  In ${\cal G}^A(x)$, $n(x)$ is the set
of links one of whose endpoints is the site $x$ and with orientation
toward $x$.  Also, we have considered the system in the presence of an
array of $K$ classical quarks which transform under representations
$R_i$ and are situated at lattice sites $x_i$.

The gauge constraint commutes with the Hamiltonian and generates
the gauge transformation,
\begin{equation}
U_\l\rightarrow U^g_\l\equiv g(x)U_\l g^{\dagger}(y) ~~,~~
E_\l\rightarrow E^g_\l\equiv g(x)E_l g^{\dagger}(x)
\end{equation}
where $g(x)$ is a unitary matrix in the fundamental representation of
the gauge group and $x$ and $y$ are the endpoints of the oriented link
$\l$: $\delta \l = [y]-[x]$.  It is useful to consider the
Schr\"odinger representation where quantum states are functions of
the gauge fields and the action of the electric field on these states
is defined algebraically.  In this representation, the gauge
constraint implies that physical states transform as
\begin{equation}
\psi_{a_1\dots a_K}[U]=g^{R_1}_{a_1b_1}(x_1)
\ldots g^{R_K}_{a_Kb_K}(x_K)
~\psi_{b_1\dots b_K}[U^g]
\label{gauge}
\end{equation}
Since the Hamiltonian is gauge invariant, the eigenstates of the
Hamiltonian carry a representation of the gauge group.  The physical
states are those which transform like (\ref{gauge}).

In the continuum limit, the spatial gauge field $\vec A$ is obtained
from $U_\l={\cal P}e^{i\int_x^y \vec A\cdot d\vec l}\approx
1+ia\vec\l\cdot\vec A+\ldots$ where $a$ is the lattice spacing and
$\vec\l$ is a unit vector in the direction of the link $\l$.  Also,
the continuum electric field operator is defined as
$E^A_\l=a^{D-1}\vec\l\cdot\vec E^A$.  In $D>1$, the magnetic energy
reduces to the magnetic field squared.  The Hamiltonian (\ref{ham}),
gauge constraint (\ref{gl}) and operator algebra (\ref{commutator})
reduce to those of continuum Yang-Mills theory,
\begin{equation}
H=\int d^Dx\left( \frac{e^2}{2}(\vec E^A)^2+\frac{1}{2e^2}(\vec B^A)^2
\right)
\end{equation}
\begin{equation}
{\cal G}^A(x)~=~(\vec D\cdot \vec E)^A
+\sum_{i=1}^K~T^A_{R_i}~\delta(x-x_i)~\sim ~0
\end{equation}
and
\begin{equation}
\left[ A_i(x),A_j(y)\right]=0~~,~~
\left[E_i(x),E_j(y)\right]=0~~,~~
\left[ E^A_i(x),A^B_j(y)\right]=-i\delta^{AB}\delta_{ij}\delta(x-y)
\end{equation}
respectively.  On the lattice, energies are measured in units of the
lattice spacing.  It is only the very low-lying states of the
dimensionless lattice Hamiltonian, with energy of order $a\cdot
\epsilon$ which have finite energy, $\epsilon$, in the continuum
limit, $a\rightarrow 0$.

\subsection{Partition function}

The partition function at temperature $T$ is the trace of the Gibbs
distribution $e^{-H/T}$ over gauge invariant physical states.  It is
convenient to take this trace in the ``coordinate'' representation.
The complete set of states is obtained by taking the direct product of
the ``position'' eigenstates of the unitary matrices
\begin{equation}\label{allstates}
\left|U\right> \equiv \prod_\l \left|U_\l \right>
\end{equation}
for each link $\l$, together with some basis elements, $e_{a_1}\dots
e_{a_K}$, which carry the representations $R_1\ldots R_K$ of the gauge
group. The states $\left|U_\l \right>$ are normalized by
\begin{equation}\label{normuu}
\left<U_\l \right.|\left.U'_\l \right>=\delta (U_\l,U'_\l),
\end{equation}
where $\delta (U_\l,U'_\l)$ is invariant $\delta $--function on the group
manifold. The states \rf{allstates} do not satisfy Gauss law constraint
\rf{gl}.  The projection on the physical subspace can be realized by
gauge transforming the
states at one side of the trace and subsequently integrating over all
gauge transforms.  The result is
\begin{equation}
Z[x_i,R_i,T]=\int\prod_\l[dU_\l]\prod_x[dg(x)]~
\left< U\right| e^{-H/T} \left| U^g \right>~{\rm Tr}g^{R_1}(x_1)
\ldots {\rm Tr}g^{R_K}(x_K)
\label{partfunc}
\end{equation}
where $[dU_\l]$ and $[dg(x)]$ are invariant Haar measures.  We
consider the case where all external quarks are in the adjoint
representation. In the adjoint representation, the trace can be taken
as the modulus squared trace of the fundamental representation group
element,
\begin{equation}
{\rm Tr} g^{\rm adj}(x)~=~ \left| {\rm Tr}~g(x)\right|^2-1
\end{equation}
We then multiply by the K'th power of the fugacity, $\lambda^K$,
multiply by $1/K!$, sum over positions, $x_1,\ldots,x_K$ and sum over
$K$.  This produces the effective theory
\begin{equation}
Z[\lambda,T]=\int\prod_\l[dU_\l]\prod_x[dg(x)]
e^{-S_{\rm eff}[U,g]}~
\label{sigma}
\end{equation}
where the effective action is
\begin{equation}
e^{-S_{\rm eff}[U,g]}~\equiv~\left<U\right|e^{-H/T}\left| U^g\right>
e^{\sum_x\lambda(\left|{\rm Tr}g(x)\right|^2-1)}
\label{kernel}
\end{equation}
This effective action has gauge symmetry,
\begin{equation}
S_{\rm eff}[ hUh^{\dagger}, hgh^{\dagger}]~=~S_{\rm eff}[U,g]
\label{gauge1}
\end{equation}
Where $h$ is an element of the gauge group.  It also has a global
symmetry
\begin{equation}
S_{\rm eff}[U, zg]~=~ S_{\rm eff}[U,g]
\label{center}
\end{equation}
where $z$ is a constant element of the center of the gauge group.
This latter symmetry is related to confinement~\cite{pol} - ~\cite{w}.
For the gauge group SU(N), the center is $Z_N$ and it is referred to
as $Z_N$-symmetry.  For U(N), the center is U(1).

When this symmetry is realized faithfully, the theory is confining.
When it is spontaneously broken, the system is in the de-confined
phase.  An order parameter for this symmetry is the Polyakov loop
operator\footnote{In the spacetime path integral formulation of finite
temperature gauge theory \cite{gpy} $A_0(\tau,x)$ is a Lagrange
multiplier field which enforces the gauge constraint.  The Euclidean
action is $$ S=\int_0^{1/T}d\tau\left( \sum_l E_\l
U_\l^{\dagger}\frac{d}{d\tau}U_\l -H[U,E]+i\sum_x A_0^A(\tau,x){\cal
G}^A(\tau,x)\right)~~, $$ and the partition function is $$Z=\int
dA_0dE[dU]e^{-S[A_0,E,U]} $$ with periodic boundary conditions in
Euclidean time.  In the strong coupling limit, this model can be
reduced to (\ref{partfunc}) by integrating the time dependent modes
explicitly.  The dynamical variable $g(x)$ is the holonomy group
element for transport of a fundamental representation charge around
the periodic Euclidean time, $$ g(x)~=~ {\cal
P}\exp\left(i\int_0^{1/T}d\tau A_0(\tau,x)\right)~~.$$ The Polyakov
loop operator is the trace of this holonomy element $ P(x)~=~{\rm
Tr}g(x)$.  } which is the trace of the group element,
\begin{equation}
P(x)~\equiv~ {\rm Tr} g(x)
\end{equation}
where $g(x)$ is in the fundamental representation of the gauge group.
It is gauge invariant and, under (\ref{center}), it transforms as
\begin{equation}
P(x)~\rightarrow~z~P(x)
\end{equation}
The quantity
\begin{equation}\label{self}
F(x,R,\lambda,T)~=~ -T\ln
\left< P(x)\right>
\end{equation}
is the free energy which is necessary to insert a classical source in
the fundamental representation into the system.  In the confining
phase, the expectation value vanishes and this free energy is
infinite.  In a de-confined phase the expectation value is non-zero
and the free energy is finite.

The two-point correlators of Polyakov loops are related to the
 free energy of the quark--antiquark pair inserted in the vacuum.
Assuming that the
cluster property holds, the two--point correlator behaves at
 $|x-y|\rightarrow \infty$ as
\begin{equation}\label{twopoint}
\left<P(x)\,P^{\dagger}(y) \right>
\longrightarrow\left| \left<P(x) \right>\right|^2
+\frac{{\rm const}}{|x-y|^{(D-1)/2}}\e^{-M_1|x-y|}+\ldots,
\end{equation}
where $M_1$ is the mass of the lowest excitation with appropriate
quantum numbers.  The two--point correlator goes to zero if the
Polyakov loop expectation value vanishes and goes to a constant
otherwise. In the deconfining phase, the quark--antiquark potential
can be defined by subtracting the self--energies \rf{self} from the
logarithm of the correlator \rf{twopoint}:
\begin{eqnarray}\label{potd}
V(x,y)&=&-T\ln\left(\left<P(x)~P^{\dagger}(y) \right>\right)
 +T\ln\langle P(x)\rangle
 +T\ln\langle P^{\dagger}(y) \rangle\nonumber\\
&\sim&
-\frac{{\rm const}}{|x-y|^{(D-1)/2}}\e^{-M_1|x-y|}.
\end{eqnarray}
In the confining phase the self--energies are infinite, but the potential
still can be defined. It grows linearly with distance:
\begin{equation}\label{potc}
V(x,y)=-T\ln\left(\left<P(x)~P^{\dagger}(y) \right>\right)
\sim \sigma |x-y|,
\end{equation}
where the string tension is
\begin{equation}\label{stringtension}
\sigma =TM_1.
\end{equation}
For the models that
we
shall consider, the mass $M_1$ will be calculated exactly in
Sec.~\ref{1d}~and~\ref{anyd}.

\subsection{Adjoint non-Abelian Coulomb gas}

We shall treat the effective action (\ref{kernel}) exactly in $D=1$
where there is no magnetic term in the Hamiltonian.  In $D>1$, we
  consider the strong coupling limit where
\begin{equation}
e^2\rightarrow\infty~~,~~ T\rightarrow\infty
~~,~~ \gamma=e^2 N/2T~{\rm finite}
\end{equation}
(We shall eventually also consider the limit where
$N\rightarrow\infty$ with $\gamma$ finite.)  Here $e^2$ and $T$ (as
well as the lattice Hamiltonian H) are dimensionless quantities which
will eventually get their engineering dimensions from factors of the
lattice spacing.  In this limit, the magnetic term in the Hamiltonian
is ignored, so that the Hamiltonian is given by the electric term only
and the Boltzmann weight \rf{kernel} factorizes on the product of the
heat kernels
\begin{equation}\label{defk}
K\left(g,h\vert\tau\right)=\left< h\left| e^{-\tau\Delta/N}\right|
g\right> \end{equation} for the Laplacian
\begin{equation}
\Delta=\sum_A \left( E^A\right)^2
\end{equation}
on the group manifold with ``time'' $\gamma=e^2N/2T$.  In this limit,
the magnetic interactions of the external quarks as well as the
magnetic self-interactions of gluons are absent.  It isolates the
electric interactions and self-interactions.

In this strong coupling limit, the partition function has the form
\begin{equation}\label{defz}
Z~=~\int\prod_x[dg(x)]\prod_\l[dU_\l]~ \prod_\l ~K\left( U_\l, gU_\l
g^{\dagger}\left|\gamma\right.\right)~ e^{\sum_x\lambda(\left|{\rm
Tr}g(x)\right|^2-1)}
\end{equation}
It is this model which we call the adjoint non-Abelian Coulomb gas.

\subsubsection{Lattice string statistical mechanics}

In strong coupling limit, the effective Hamiltonian is
\begin{equation}\label{ho}
H_0=\sum_{\l,A} \frac{e^2}{2} (E^A_\l)^2
\end{equation}
The local moments of the electric field distribution are conserved,
\begin{equation}
\left[ E^A(x), H_0 \right]=0
\end{equation}
and distribution of electric fields is frozen in time. The eigenstates
of the Hamiltonian \rf{ho} can be constructed by acting with link
variables $U_l$ on the vacuum. On a given link, we begin with the
singlet state,
\begin{equation}
E^A_\l \left| 0\right>~=~0
\end{equation}
and the action of $E^A_\l$ on excited states is defined
combinatorially,
\begin{equation}
E^A_\l~(U_\l)_{ab} \left| 0\right>~=~(T^AU_\l)_{ab}\left| 0\right>
\end{equation}
\begin{equation}
E^A_\l~ (U_\l)_{ab}(U_\l)_{cd}\left|0\right>= ~
(T^AU_\l)_{ab}(U_\l)_{cd}\left|0\right>+
(U_\l)_{ab}(T^AU_\l)_{cd}\left|0\right>
\end{equation}
and so on. The gauge invariant states contain closed loops of electric
flux and lines of electric flux, appropriate numbers of which end at
the quarks.  Closed lines of flux are created by the traces of the
spatial Wilson loop operators in different representations of the
gauge group, for the closed curve $\Gamma$,
\begin{equation}
W_R[\Gamma]={\rm Tr}\prod_{\l\in\Gamma} U^R_\l
\end{equation}
For a non-intersecting array of electric flux strings, the
energy is proportional to the total length of all strings,
\begin{equation}
{\it E}= \frac{e^2}{2}C_2\sum_{\Gamma }L[\Gamma]
\end{equation}
\begin{figure}[tbp]
\unitlength 0.70mm
\linethickness{0.8pt}
\begin{picture}(105.00,10)(-37,38)
\put(70.00,40.00){\vector(1,0){0.2}}
\bezier{120}(40.00,40.00)(45.00,40.00)(70.00,40.00)
\put(70.00,45.00){\vector(1,0){0.2}}
\bezier{120}(40.00,45.00)(45.00,45.00)(70.00,45.00)
\bezier{40}(70.00,45.00)(80.00,45.00)(80.00,45.00)
\bezier{40}(70.00,40.00)(80.00,40.00)(80.00,40.00)
\bezier{40}(130.00,45.00)(140.00,45.00)(140.00,45.00)
\bezier{40}(130.00,40.00)(140.00,40.00)(140.00,40.00)
\bezier{64}(100.04,44.98)(111.65,44.77)(115.00,42.47)
\put(129.96,39.96){\vector(1,0){0.2}}
\bezier{60}(115.00,42.37)(118.66,40.38)(129.96,39.96)
\bezier{60}(100.00,40.01)(109.09,39.32)(114.95,41.39)
\put(130.03,44.95){\vector(1,0){0.2}}
\bezier{56}(117.02,42.77)(121.86,45.99)(130.03,44.95)
\put(25.00,42.33){\makebox(0,0)[cc]{$|\Psi_{\pm}\rangle~=$}}
\put(90.00,42.33){\makebox(0,0)[cc]{$\pm$}}
\end{picture}
\caption {Graphical representation of the states
 $\left|\Psi_{\pm}\right>$}
\label{picpsi}
\end{figure}

Strings interact when they have common links.  For example, we
consider an array of strings where two intersecting strings share a
common link.  In this case, the eigenstates of the Hamiltonian are
linear combinations of the uncrossed and crossed strings,
fig.~\ref{picpsi}:
\begin{equation}\label{defpsi}
\left|\Psi_{\pm}\right>=\left(
U_{ab}U_{cd}\pm U_{ad}U_{cb}\right)\left|0\right>.
\end{equation}
The contribution of the states $\left|\Psi_{\pm}\right>$ to the energy
of the string configuration is equal to $e^2(N-1)(N+2)/N$ and
$e^2(N+1)(N-2)/N$, respectively.  In the large $N$ limit, these states
are degenerate and diagonalization of the Hamiltonian does not mix the
crossed and uncrossed strings.  Therefore, the interaction vanishes in
the large N limit.  Thus, in the large $N$ limit, the statistical
mechanics of the strong coupling Yang-Mills theory is equivalent to a
statistical model of non-interacting lattice strings for which the
partition sum can be written
\begin{equation} Z=\sum_{\Gamma} e^{-C_2
\gamma L[\Gamma]/2N}~=~\sum_L n(L)e^{-C_2\gamma L/2N}.
\end{equation}
Such a system is expected to have a percolation transition when the
entropy of configurations of strings overtakes the energy \cite{BR}.
For large curves, the number of strings with a given length $L$ grows
like
\begin{equation} n(L)~\sim~{\rm const.}~ (2D-1)^L
\end{equation}
The critical temperature is given by
\begin{equation} \gamma_{\rm
crit.}~=~ 2\ln(2D-1) \end{equation} This is the result which was found
in \cite{Zar,Boul} by the direct solution of the model.

When there are adjoint quarks present, quantum states still contain
closed loops of electric flux.  In addition, it is necessary that each
quark absorbs and emits one line of flux.  Increasing the density of
quarks decreases the phase transition temperature somewhat.

\subsubsection{Continuum limit}

In the continuum limit, the trace which is taken to obtain the
partition function can be taken over a complete set of eigenstates of
the gauge field operator, twisted by the gauge
transformation~\cite{gsst}, $\left< A\right| e^{-H/T}\left|
A^g\right>$, where $A^g=g(A+id)g^{\dagger}$.  Formally, we can
consider the strong coupling limit and retain only the electric term
in the Hamiltonian:
\begin{eqnarray}
\left<A\right| e^{-\frac{e^2}{2T}\int(E^A)^2}\left|A^g\right>
&=&{\rm const.}~\exp\left(-\frac{T}{e^2}\int d^Dx\,
 {\rm Tr}(A-A^g)^2\right)
\nonumber\\
&=&{\rm const.}~\exp\left(-\frac{T}{e^2}\int d^Dx\,
 {\rm Tr}\left| Dg(x)\right|^2
\right)
\end{eqnarray}
where $Dg=\nabla g+i\left[A,g\right]$.  The partition function is
\begin{equation}
Z~=~\int [dA][dg] ~e^{-\int d^Dx\,\left[\frac{N}{2\gamma}{\rm
Tr}\left|Dg(x)\right|^2-\lambda(|{\rm Tr}g(x)|^2-1)\right]}
\label{chiral}
\end{equation}
The effective action is that of a gauged principal chiral model in $D$
dimensions. The continuum treatment of the strong coupling limit,
however, is valid only at $D=1$, where the magnetic term in the
Hamiltonian is absent from the beginning; this model is considered in
the next section. At $D>1$ the field theory defined by \rf{chiral} is
nonrenormalizable unless the kinetic term for the gauge fields is
added.

\vskip 0.5truein

The mean density of quarks is obtained from the partition function as
\begin{equation}\label{chdens}
\frac{\langle n\rangle}{V}=\frac{1}{V}
\lambda \frac{\partial }{\partial \lambda }\ln Z\left[\lambda,T\right]
=\lambda \langle{\rm Tr} g^{\dagger}{\rm Tr} g-1\rangle
\rightarrow \lambda
\left|\left<{\rm Tr} g\right>\right|^2 +O(N^0) ~~({\rm as}~N\rightarrow \infty)
\end{equation}
where we have assumed that the expectation value is translation
invariant.  The last equality follows from factorization of invariant
correlators in the large $N$ limit. From eq. \rf{chdens} we see that,
in the large $N$ limit, the Coulomb gas in the confining phase where
$\left|\left< {\rm Tr}~g(x)
\right>\right|=\left|\left<P(x)\right>\right| = 0$ is dilute
with density is of order one.  This is a result of the fact that the
number of species of glue-balls and hadrons does not grow in the large
$N$ limit.  Above the deconfining temperature, where $\left|\left<
P(x)\right>\right|\neq 0$ there are $N^2-1$ gluon and $N^2$ quark
degrees of freedom and the density is of order $N^2$.  A similar
consideration applies to the free energy.  The free energy in the
confining phase should grow like the number of degrees of freedom and
be ${\cal O}(N^0)$ in the large $N$ limit.  In the de-confined phase
the free energy should be of order $N^2$. For $N$ infinite, this
implies that a phase transition between the two phases would have
infinite latent heat.  However, one should consider the large $N$
limit as analogous to the infinite volume limit in statistical
mechanics where, strictly speaking, there is no phase transition in
finite volume, but the analysis in infinite volume is to a good
approximation accurate in the physical finite system too.

This will be particularly relevant to the case of $D=1$, where, if $N$
is finite, the dimensionality of the system is too low to have
spontaneous symmetry breaking.  It has local interactions and is
effectively a one-dimensional system. This implies that, when $N$ is
finite, it is always in the confining phase with unbroken center
symmetry.  In the infinite $N$ limit, there can be a phase transition
and a phase with broken symmetry.  However, the latter limit should
describe the physical behavior of the system accurately even when $N$
is finite if the ``finite size'' corrections of order $1/N^2$ are
small.

Even this case is subtle, since the effective symmetry in the infinite
$N$ limit is U(1), a continuous symmetry and the effective dimension
is two.  The small $\gamma$ phase can be ordered only because the
action is non-local in index space, similar to an infinite-ranged spin
model.

In Section 2 we shall analyze the large N limit of the one dimensional
model directly in the continuum.  We find an exact solution of the
continuum model and show explicitly that it has a first order
confinement-deconfinement phase transition which occurs at a critical
line in the $\gamma-\lambda$ plane.  In Section 3 we discuss the
strong coupling lattice model in any dimensions.  We find an exact
solution in the confined phase and an approximate solution of the
deconfined phase.  As on one dimension, there is a first order
confinement-deconfinement transition.  In Section 4, we discuss the
results.

\newsection{Large $N$ limit of one--dimensional model}\label{1d}

It is possible to analyze the large $N$ limit of the D=1 model
directly in the continuum limit. We consider the gauge group $U(N)$
which has center $U(1)$, or $SU(N)$ with center $Z_N$ -- there is no
difference in the large $N$ limit.  In the gauged principal chiral
model (\ref{chiral}), since the space is an open line, we can choose
the gauge $A=0$.  The partition function for the resulting model,
\begin{equation}\label{z1d}
Z=\int\,[Dg]\,\e^{-\int\,dx\,\left[\frac{N}{2\gamma}
\tr(g^{\prime})^{\dagger}(g^{\prime})-\lambda \tr g^{\dagger} \tr g\right]}
\end{equation}
is equivalent to that of unitary matrix quantum mechanics (where $x$
is imaginary time) and can be solved in the large $N$ limit by the methods
of collective field theory ~\cite{JS,Wadia}.  The method is essentially
based on the relation between matrix quantum mechanics and
nonrelativistic fermions ~\cite{BIPZ}.

The dynamical variables in (\ref{z1d}) are the phases $\alpha_k(x)$ of
the eigenvalues $\e^{i\alpha_k(x)}$ of the Polyakov loop variables
$g(x)$.  They can be interpreted as coordinates of fermions which live
on a circle.  In the large $N$ limit, the statistical integral in
(\ref{z1d}) is dominated by a single distribution of eigenvalues.  We
introduce the eigenvalue density
\begin{equation}
\label{density}
\rho(\theta ,x)=\frac{1}{N}\sum_{k=1}^{N}\,\delta
(\theta -\alpha_{k}(x)),
\end{equation}
which is a periodic function which is normalized to unity on the
interval $(-\pi ,\pi )$. It has the mode expansion
\begin{equation}
\rho(\theta,x)=\frac{1}{2\pi}\left( 1+ \sum_{n\neq 0}c_n(x)e^{in\theta}
\right)~~~,~~c_n^*=c_{-n}
\label{densitymodes}
\end{equation}
In a truly confining phase, all moments of the eigenvalue distribution
vanish, $\rho_{\rm conf.}=1/2\pi$.  If any of the other moments are
non-zero, there are Polyakov loop operators which have non-zero
expectation values, if $c_n(x)\neq 0$ then $\left< {\rm
Tr}~g^n(x)~\right>\neq 0$.  Thus, all combinations of $c_n(x)\neq 0 $
characterize all de-confined phases.

In the large $N$ limit the eigenvalue density obeys a classical,
saddle-point equation which can be deduced from canonical analysis of
the collective field theory Hamiltonian\cite{JS}, \cite{JSD}:
\begin{equation}
\label{jsham}
H=\int d\theta
\,\left[\frac{\gamma }{2}\rho (\theta ) \left(\frac{\partial
\Pi}{\partial \theta }\right)^2 +\frac{\pi ^2\gamma }{6}\rho ^3(\theta
)\right] -\lambda \left|\int d\theta \,\rho (\theta )\e^{i\theta
}\right|^2 -\frac{\gamma}{24}
\end{equation}
with subsequent Wick rotation to an imaginary time.
Here $\Pi(\theta,x )$ is the variable which is the canonical conjugate
to $\rho (\theta,x )$, so that the Poisson bracket is
\begin{equation}
\left\{ \rho(\theta,x),\Pi(\theta',x)\right\}=\delta(\theta-\theta')
\end{equation}
(here $x$ is the time variable) and $v(\theta )={\partial\Pi}/
{\partial\theta }$ is the velocity of the Fermi fluid. The equations
of motion, following from \rf{jsham}, read, after the change
$x\rightarrow ix$, $v\rightarrow -iv$, as follows:
\begin{equation}
\label{cont}
\frac{\partial \rho }{\partial x}+\gamma \frac{\partial }{\partial
\theta }(\rho v)=0,
\end{equation}
\begin{equation}
\label{euler}
\frac{\partial v}{\partial x}+\gamma
v\frac{\partial v}{\partial \theta } -\pi
^2\gamma \rho \frac{\partial \rho }{\partial \theta } +2\lambda
\impart\left(\e^{-i\theta }c_1(x)\right)=0, \end{equation}
where
\begin{equation}\label{defc} c_1(x)=\int_{-\pi }^{\pi }\,d\theta \,\rho
(\theta ,x)\e^{i\theta }.
\end{equation}
It is expected that the solution of these equations corresponding to
the physical vacuum is a constant $\rho_0(\theta )$.

One may expect that, at least at sufficiently low temperature or,
equivalently, at sufficiently large $\gamma$, the system is in the
confining phase with unbroken center group symmetry. This symmetry
acts on $\rho (\theta ,x)$ by translation $\theta \rightarrow\theta
+\theta _0$, so the only symmetric solution is $\rho_{\rm
conf.}(\theta )=\frac{1}{2\pi }$. It always satisfies the equations of
motion, as $c_1(x)$, defined by eq. \rf{defc}, is equal to zero when
$\rho=\rho_{\rm conf.}$.

However, this solution, which is an exact solution of (\ref{cont}) and
(\ref{euler}), is stable to small fluctuations only if $\gamma$ is
large enough.  It becomes unstable for $\gamma <\gamma _c(\lambda )$.
To see this, it is necessary to analyze the spectrum of fluctuations
in the strong coupling phase.  Consider $\rho (\theta
,x)=\frac{1}{2\pi }(1+\varphi (\theta ,x))$ and consider the equations
(\ref{cont}) and (\ref{euler}) to linear order in $\varphi $:
\begin{equation}
\label{lcont}
\frac{\partial \varphi }{\partial
x}+\gamma \frac{\partial v}{\partial \theta }=0,
\end{equation}
\begin{equation}
\label{leuler}
\frac{\partial
v}{\partial x}-\frac{\gamma }{4}\,\frac{\partial \varphi } {\partial
\theta }+2\lambda \impart(c_1\e^{-i\theta })=0.
\end{equation}
Differentiating eq. \rf{lcont} with respect to $x$ and eq.
~\rf{leuler} with respect to $\theta $ and subtracting the latter,
multiplyed by $\gamma $, from the former, we obtain an equation for
the density fluctuations:
\begin{equation}
\label{linear}
\frac{\partial ^2\varphi }{\partial x^2}+\frac{\gamma
^2}{4}\frac{\partial ^2\varphi }{\partial \theta ^2}+2\lambda \gamma
\realpart(c_1\e^{-i\theta }) =0.
\end{equation}
The eigenmodes are the Fourier harmonics of $\varphi (\theta ,x)$:
\begin{equation}
\varphi (\theta ,x)=\sum_{n\neq 0}c_n(x)\e^{in\theta }
\end{equation}
where now $c_n$ are infinitesimal:
\begin{equation}\label{spec}
c_n''-\left[\frac{\gamma ^2n^2}{4}-\lambda \gamma (
\delta_{n,1}+\delta_{n,-1})\right] c_n=0~~,~~n\neq 0.
\end{equation}
We obtain the following spectrum of excitations:
\begin{equation}\label{massn}
M_n^2=\frac{\gamma ^2n^2}{4}-\lambda \gamma \delta _{n\,1},~~n=1,2,\ldots
\end{equation}

At $\gamma =\gamma _c(\lambda )$, where \begin{equation}\label{gc}
\gamma _c(\lambda )=4\lambda , \end{equation} the lowest eigenvalue
($n=1$) goes to zero. For smaller $\gamma $ this eigenvalue is
negative and leads to the instability of the strong coupling solution
where $c_1$ is the first mode to become unstable. However, for reasons
which will become clear once we consider the weak coupling phase,
$\gamma _c(\lambda )$ should not be identified with the point of the
deconfining phase transition.

The contribution of fluctuations to the free energy is given by
\begin{equation}
\delta F/V= \pi T
\int_{-\infty}^\infty \frac{dk}{2\pi}\left(\sum_n \ln
\left(k^2 +\frac{\gamma^2n^2}{4}\right) +\ln\left(\frac{ k^2+\gamma^2/4
-\lambda\gamma}{k^2+\gamma^2/4}\right)\right)
\end{equation}
the singular part of which, near the critical line is $\delta F/V\sim
\frac{\pi^2\gamma T}{2}\left(\sqrt{1-4\lambda/\gamma} -1\right)$.

The deconfining solution can be obtained by integration of eq.
\rf{euler} at $v=0$. The density $\rho _0(\theta )$ can always be
chosen to be an even function of $\theta $.  Thus $c_1$ is real, and
one finds from eq. \rf{euler}:
\begin{equation}\label{wcs}
\rho_0(\theta )=\frac{1}{\pi }\sqrt{\frac{2}{\gamma }\Big(E +2\lambda c_1
\cos\theta\Big)}.
\end{equation}
The Fermi energy $E$ and the constant $c_1$ are to be determined from
the normalization condition and eq. \rf{defc}:
\begin{equation}\label{norm}
\int_{-\tmax}^{\tmax}\,\frac{d\theta }{\pi }\, \sqrt{\frac{2}{\gamma
}\Big(E+2\lambda c_1 \cos\theta \Big)} =1,
\end{equation}
\begin{equation}\label{c}
\int_{-\tmax}^{\tmax}\,\frac{d\theta }{\pi
}\,\cos\theta\, \sqrt{\frac{2}{\gamma }\Big(E+2\lambda c_1 \cos\theta
\Big)} =c_1,
\end{equation}
\begin{equation}\label{tmax}
\tmax=\pi
-\arccos\frac{E}{2\lambda c_1}.
\end{equation}
It follows from these equations that $\tmax$ tends to zero at $\gamma
\rightarrow 0$ and grows with the increase of $\gamma $. Eventually it
reaches $\pi $, where the weak coupling phase terminates, because the
eigenvalue distribution begins to overlap with itself due to $2\pi
$--periodicity.  At the critical point $E_*=2\lambda c_{1*}$, the
integrals in \rf{norm} and \rf{c} can be done explicitly, and we find
that $c_{1*}=\frac{1}{3}$ and
\begin{equation}\label{g*}
\gamma_*(\lambda )=\frac{128}{3\pi ^2}\,\lambda=4.323\,\lambda .
\end{equation}

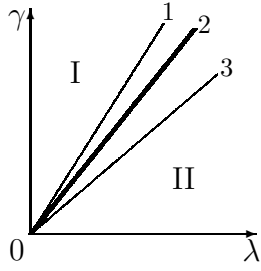
\begin{figure}[tbp]
\unitlength 0.3mm
\linethickness{0.4pt}
\begin{picture}(120.00,120.00)(-200,0)
\put(120.00,20.00){\vector(1,0){0.2}}
\put(20.00,20.00){\line(1,0){100.00}}
\put(20.00,120.00){\vector(0,1){0.2}}
\put(20.00,20.00){\line(0,1){100.00}}
\multiput(20.00,20.00)(0.12,0.19){492}{\line(0,1){0.19}}
\multiput(20.00,20.00)(0.14,0.12){592}{\line(1,0){0.14}}
\linethickness{1.6pt}
\multiput(20.00,20.00)(0.12,0.15){609}{\line(0,1){0.15}}
\linethickness{0.4pt}
\put(14.00,115.00){\makebox(0,0)[cc]{$\gamma$}}
\put(118.00,13.00){\makebox(0,0)[cc]{$\lambda$}}
\put(14.00,13.00){\makebox(0,0)[cc]{$0$}}
\put(40.00,92.00){\makebox(0,0)[cc]{I}}
\put(88.00,45.00){\makebox(0,0)[cc]{II}}
\put(81.00,119.00){\makebox(0,0)[cc]{{\footnotesize 1}}}
\put(97.00,113.00){\makebox(0,0)[cc]{{\footnotesize 2}}}
\put(107.00,94.00){\makebox(0,0)[cc]{{\footnotesize 3}}}
\end{picture}
\caption {
 The large $N$ phase diagram of the one--dimensional model. I --
strong coupling (confining) phase, II -- weak coupling (deconfining)
phase; 1 -- line on which the weak coupling phase terminates:
$\gamma=\gamma_*(\lambda)$, 2 -- line of the first--order phase
transition: $\gamma =\gamma _0(\lambda )$, 3 -- line of the
instability of the strong coupling phase: $\gamma=\gamma_c(\lambda)$}
\label{1} \end{figure}

At this critical line, the weak coupling phase is unstable.  It is
there that the translation zero mode of the linearization of the
equations\rf{cont}-\rm{defc}, $\partial\rho/\partial\theta$ becomes
normalizable and the resulting fluctuations restore the translation
invariance in $\theta$, i.e. the center symmetry.  In the phase with
density given by \rm{wcs} this zero mode is not normalizable and
therefore it is ineffective in restoring the symmetry.

We obtain the following picture of the deconfining phase transition
(fig.  \ref{1}). The weak and strong coupling phases can coexist,
because $\gamma _c(\lambda )<\gamma _*(\lambda )$, although the
region, where both phases are stable is very narrow, since $\gamma
_c(\lambda )$ and $\gamma _*(\lambda )$ are numerically closed to each
other. The phase transition is of the first order and takes place at
some $\gamma_0(\lambda ) $ between $\gamma _c(\lambda )$ and $\gamma
_*(\lambda )$ (fig.  \ref{1}). The line of phase transitions is
defined as that line where the free energies of the both phases are
equal to each other.  Substituting $\rho _0(\theta )$ into
eq.\rf{jsham} one can find the free energy per unit volume, to leading
order in the large $N$ limit:
\begin{equation}\label{free}\frac{ F}{VN^2} =\left\{\begin{array}{ll}
0,&{\rm in~confining~phase},\\
\frac{1}{3}E-\frac{1}{3}\lambda c_1^2 -\frac{\gamma}{24},&{\rm
in~deconfining~phase}.\\ \end{array}\right.  \end{equation} The
equations determining the critical point can be solved numerically,
the result is:
\begin{equation}\label{phtrp}
 \gamma _0(\lambda )=4.219\,\lambda .
\end{equation}

\newsection{Strong coupling lattice model in any dimension}\label{anyd}

The model \rf{defz} is an unitary matrix analog of the Kazakov--Migdal
model \cite{KM}, which can be treated at large $N$ by the saddle point
methods \cite{Mig}--\cite{Mat}.  Similar methods have been applied to
the model \rf{defz} with $\lambda =0$
\cite{Zar}--\cite{BCDMP}. Here we generalize the consideration of
\cite{Zar} to the case of $\lambda \geq 0$.

To begin, we choose the gauge in which the Polyakov loops are
diagonal: $g(x)=\diag\left(\e^{i\alpha _k(x)}\right)$. Then,
integrating over the gauge fields $U_l$ we obtain an effective
action for the eigenvalues $\alpha _k(x)$:
\begin{eqnarray}\label{seff} S_{{\rm eff}}=&-&\sum_{x}\left[\lambda
\sum_{kj} \e^{i\alpha _k(x)-i\alpha _j(x)}
+\sum_{k<j}\ln\sin^2\frac{\alpha _k(x)-\alpha _j(x)}{2}\right.
\nonumber\\ &+&\left.\frac{1}{2}\sum_{\mu =-D}^{D}\ln I(\alpha (x),
\alpha (x+\mu )\left|\gamma \right.)\right], \end{eqnarray} where the
second term comes from the Faddeev--Popov determinant and $I(\alpha
,\alpha '\left|\tau \right.)$ is a one--link integral:
\begin{equation}\label{defi} I(\alpha ,\alpha '\left|\tau
\right.)=\int\,DU\,K(\e^{i\alpha }, U\e^{i\alpha
'}U^{\dagger}\left|\tau \right.),  \end{equation}
and we have used the invariance properties of the heat kernel.
The link integral is calculable for any $N$ \cite{FAI} and can be
represented in the following form:  \begin{equation}\label{int} I(\alpha
,\alpha '\left|\tau \right.)={\rm const}\cdot\frac{
\det_{kj}\vartheta\left(\frac{\alpha _k-\alpha'_j}{2\pi }
\left|\frac{i\tau }{2\pi N}\right.\right)}{J(\alpha )J(\alpha ')},
~~~~J(\alpha )=\prod_{i<j}\sin\frac{\alpha _i-\alpha _j}{2},
\end{equation} where $\vartheta(z\left|\tau \right.)$ is the Riemann
theta function.

In the large $N$ limit, since the effective action \rf{seff} is of
order $N^2$ and there are $N$ degrees of freedom, the statistical sum
is dominated by the configuration which minimizes the action, i.e. the
solution of the classical equation of motion.  In terms of the
eigenvalue density \rf{density} the equation of motion reads:
\begin{eqnarray}\label{eqm} &&-2\lambda
\impart\left(c_1(x)\e^{-i\theta }\right) -\wp\int_{-\pi }^{\pi
}\,d\theta' \,\rho (\theta ',x)\cot\frac{ \theta -\theta
'}{2}\nonumber\\ &&=\sum_{\mu =-D}^{D}\frac{1}{N^2}\,\frac{\partial
}{\partial \theta } \,\frac{\delta }{\delta \rho (\theta ,x)}\,\ln
I(\alpha(x) ,\alpha (x+\mu )\left|\gamma \right.).
\end{eqnarray}
where
\begin{equation}\label{defc'}
c_1(x)=\int_{-\pi }^{\pi }\,d\theta\,\rho (\theta ,x)\e^{i\theta }.
\end{equation}
The large $N$ limit of the one--link integral \rf{defi} can be
considered analogously to it's Hermitean--matrix counterpart along the
lines of \cite{IZ,Mat}.  The method again is based on the
correspondence between matrix quantum mechanics -- cf. eqs.  \rf{defk}
and \rf{defi} (the integration in
\rf{defi} acts as a projection on the singlet states) -- and the
quantum mechanics of the free nonrelativistic fermions -- note that
the numerator in eq. \rf{int} has a form of the Slater determinant. In
the large $N$ limit the fermions behave semiclassically and the
problem reduces to the equations of motion of the collective field
theory: \begin{equation}\label{cont'} \frac{\partial \sigma }{\partial
\tau }+\frac{\partial }{\partial \theta }(\sigma s)=0, \end{equation}
\begin{equation}\label{euler'} \frac{\partial s}{\partial \tau
}+s\frac{\partial s}{\partial \theta } -\pi ^2\sigma \frac{\partial
\sigma }{\partial \theta } =0, \end{equation} which should be solved
on each link of the lattice with the following boundary conditions:
\begin{equation}\label{bound1} \sigma (\theta ,0;x,\mu )=\rho (\theta
,x), \end{equation} \begin{equation}\label{bound2} \sigma (\theta
,\gamma ;x,\mu )=\rho (\theta ,x+\mu ).  \end{equation} The right hand
side of eq. \rf{eqm} can be expressed through the solution of
\rf{cont'} -- \rf{bound2} as follows: \begin{eqnarray}\label{eqm'}
&&-2\lambda \impart\left(c_1(x)\e^{-i\theta }\right)
+(D-1)\wp\int_{-\pi }^{\pi }\,d\theta' \,\rho (\theta ',x)\cot\frac{
\theta -\theta '}{2}\nonumber\\ &&=\sum_{\mu =-D}^{D}\Big(s(\theta
,0;x,\mu )-s(\theta ,\gamma ; x-\mu ,\mu )\Big).  \end{eqnarray}

 The equations \rf{defc'} -- \rf{eqm'} completely determine the large
$N$ dynamics of the model under consideration. The vacuum should be
identified with the $x$--independent solution. At $D=1$ these
equations have a continuum limit. One should introduce the lattice
spacing $a$, recover the canonical dimensions of the couplings, i.e.
rescale $\gamma \rightarrow a\gamma $ and $\lambda \rightarrow a
\lambda $, and take the limit $a\rightarrow 0$. As a result of this
procedure, one obtains eqs.  \rf{cont} -- \rf{defc} of sec. \ref{1d}
\cite{Zar}.

 The vacuum solution in the large $\gamma $ confining phase is
dictated by the center group symmetry -- $\rho _0(\theta
)=\frac{1}{2\pi }$. To find the spectrum of excitations we linearize
the equations of motion around $\rho _0(\theta )$:
\begin{equation}\label{f1} \rho (\theta ,x)=\frac{1}{2\pi
}+\frac{1}{2\pi }\sum_{n\neq 0}c_n(x) \e^{in\theta },~~~~c_n^*=c_{-n},
\end{equation} \begin{equation}\label{f2} \sigma (\theta,\tau ;x,\mu
)=\frac{1}{2\pi }+\frac{1}{2\pi }\sum_{n\neq 0}\alpha _n(\tau ;x,\mu )
\e^{in\theta },~~~~\alpha _n^*=\alpha _{-n}, \end{equation}
\begin{equation}\label{f3} s(\theta,\tau ;x,\mu )=\sum_{n\neq 0}\beta
_n(\tau ;x,\mu ) \e^{in\theta },~~~~\beta _n^*=\beta _{-n}.
\end{equation} The solution of eqs. \rf{cont'}, \rf{euler'} linearized
in $\alpha_n $ and $\beta_n $ reads: \begin{equation}\label{an} \alpha
_n(\tau ;x,\mu )=\alpha_n ^+(x,\mu )\e^{\frac{n\tau }{2}} +\alpha_n
^-(x,\mu )\e^{-\frac{n\tau }{2}}, \end{equation}
\begin{equation}\label{bn} \beta _n(\tau ;x,\mu )=\frac{i}{2}\left[
\alpha _n^+(x,\mu )\e^{\frac{n\tau }{2}} -\alpha _n^-(x,\mu
)\e^{-\frac{n\tau }{2}}\right].  \end{equation} Substituting these
equations in the boundary conditions \rf{bound1}, \rf{bound2} we
express $\alpha _n$ and $\beta _n$ in terms of $c_n$:
\begin{equation}\label{a+-n} \alpha _n^{\pm}(x,\mu
)=\mp\frac{\e^{\mp\frac{n\gamma}{2}}c_n(x) -c_n(x+\mu
)}{2\sinh\frac{n\gamma }{2}}.  \end{equation} The equality \rf{eqm'}
then gives an equation for the Fourier coefficients of the eigenvalue
density: \begin{equation}\label{linear'} \sum_{\mu
=1}^{D}\left\{c_n(x+\mu ) -2\left[\cosh\frac{n\gamma }{2}
-\left(\frac{D-1}{D}+\delta _{n\,1}\lambda \right)\sinh\frac{n\gamma
}{2}\right]c_n(x)+c_n(x-\mu )\right\}=0, \end{equation} which leads to
the following mass spectrum: \begin{equation}\label{mass}
M_n^2=2D\left(\cosh\frac{n\gamma }{2}-1\right)-2(D-1+\delta
_{n\,1}D\lambda )\,\sinh\frac{n\gamma }{2}.  \end{equation} The strong
coupling solution becomes unstable when $M_1^2$ turns to zero.  Form
eq. \rf{mass} one obtains: \begin{equation}\label{gc'} \gamma
_c(\lambda )=2\ln\frac{D+\sqrt{(D-1)^2+\lambda D [2(D-1)+\lambda
D]}}{1-\lambda D}.  \end{equation}

 We have not been able to obtain an exact solution in the weak
coupling phase, but an approximate one, valid in the limit $\gamma
\rightarrow 0$, can be found. At small $\gamma$ one can expand
$g(x)=\e^{i\sqrt{\gamma} \Phi (x)}\approx 1+i\sqrt{\gamma} \Phi (x)$,
and the partition function \rf{defz} reduces to that of the
Kazakov--Migdal model with the quadratic potential. From the point of
view of the large $N$ equations of motion this approximation
corresponds to the expansion of the left hand side of eq. \rf{eqm'} in
the powers of $\theta $.  Actually, since at small $\gamma $ the
eigenvalues are peaked about zero with the width of the distribution
of order $\gamma $, we can expand the second term in eq. \rf{eqm'},
retaining only the contributions of order $\gamma ^{-1}$ and $\gamma
^0$: \begin{equation}\label{exp} \wp\int\,d\theta' \,\rho (\theta',x
)\cot\frac{\theta -\theta '}{2} =2\wp\int\,d\theta '\,\frac{\rho
(\theta ',x)}{\theta -\theta '} -\frac{1}{6}\theta -O(\gamma ^2).
\end{equation} The first term in \rf{eqm'} is equal, with the same
accuracy, to $2\lambda \theta $, as $c_1=1+O(\gamma ^2)$. In this
approximation, we obtain the equations of motion for the
Kazakov--Migdal model with the quadratic potential, as expected, and
effective mass is equal to \begin{equation}\label{meff}
\meff2=2D+\left[2\lambda -\frac{1}{6}(D-1)\right]\gamma .
\end{equation} The vacuum solution of the latter model is known
\cite{Gr}: \begin{equation}\label{weak} \rho _0(\theta )=\frac{1}{\pi
}\sqrt{\mu-\frac{1}{4}\mu ^2\theta ^2}, \end{equation}
\begin{equation}\label{mu} \mu =\frac{\meff2(D-1)+D\sqrt{m^4_{{\rm
eff}}-4(2D-1)}}{(2D-1)\gamma }.  \end{equation}

 There are two reasons for which the weak coupling solution can
terminate.  First, it may become unstable due to the appearance of the
massless excitation in the spectrum of fluctuations around it.  For
the solution \rf{weak}, \rf{mu} of the Kazakov--Migdal model it
happens at $\meff2=2\sqrt{2D-1}$ \cite{AG}; for smaller values of
$\meff2$ this solution does not exist since $\mu $ in \rf{mu} becomes
complex. However, the analyses of the $\lambda =0$ case shows
\cite{BCDMP} that the weak coupling phase terminates before it reaches
the point of instability, because the eigenvalue distribution hits
$\pi $ and begins to overlap with itself for smaller $\gamma $. More
precise criterion for a critical point is the condition that the width
of the support of $\sigma (\theta ,\tau )$ exceeds $2\pi$ at some
intermediate $\tau $ between $0$ and $\gamma $ \cite{BCDMP}. For the
solution \rf{weak}, \rf{mu}, the last criterion is satisfied when
 \cite{BCDMP}:
\begin{equation}\label{est}
\gamma _*(\lambda )^2=\pi ^4-\frac{4\pi ^2}{\mu (\gamma _*(\lambda ),
\lambda )}.
\end{equation}
The solution \rf{weak}, \rf{mu} being an approximate one, eq.~\rf{est}
gives only an estimate for the critical coupling. In fact, it should
give an upper bound. The arguments for this are the following: All
higher terms in the expansion of the cotangent in \rf{exp} are
negative, or, equivalently, the effective potential generated due to
nonlinearity of the field $g(x)$, is upside--down. Thus the neglected
corrections can only strengthen the instability. These arguments are
rigorous for $\lambda =0$.  When $\lambda>0$, the first term in
\rf{eqm'} which is proportional to $\lambda$ has an alternating Taylor
expansion in powers of $\theta$ when $c_1$ is real.  However, the
first correction is also negative and we expect that \rf{est} gives an
upper bound for the actual value of $\gamma _*(\lambda )$ for all
$\lambda $.

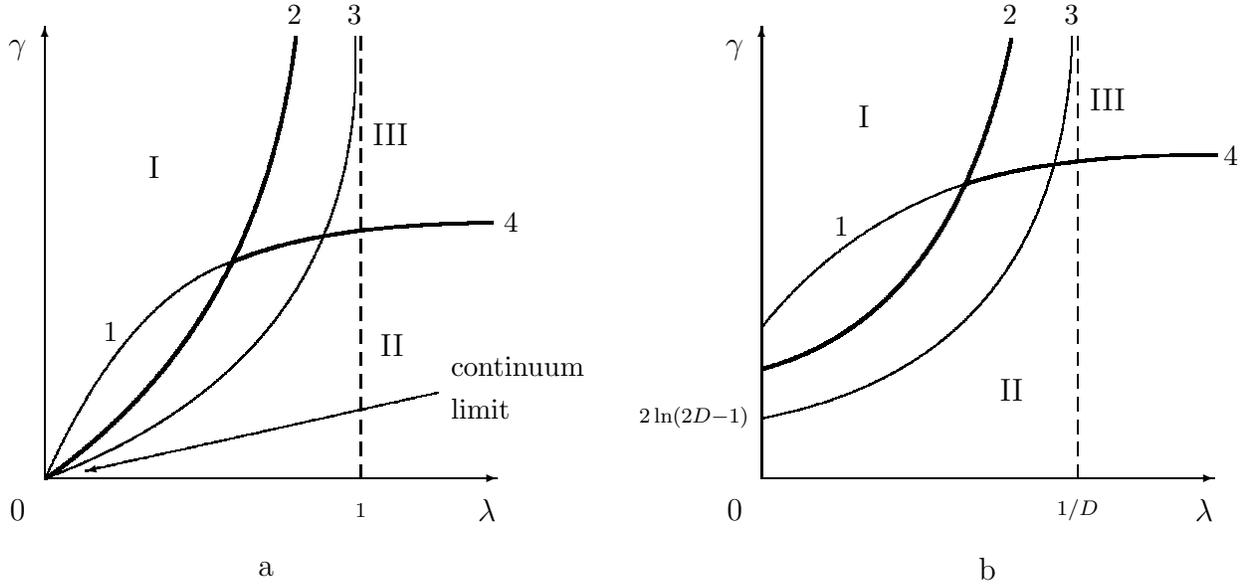
\begin{figure}[tbp]
\unitlength 0.60mm
\linethickness{0.4pt}
\begin{picture}(282.89,133.00)(13,-8)
\put(120.00,20.00){\vector(1,0){0.2}}
\put(20.00,20.00){\line(1,0){100.00}}
\put(20.00,120.00){\vector(0,1){0.2}}
\put(20.00,20.00){\line(0,1){100.00}}
\put(14.00,115.00){\makebox(0,0)[cc]{$\gamma$}}
\put(118.00,13.00){\makebox(0,0)[cc]{$\lambda$}}
\put(14.00,13.00){\makebox(0,0)[cc]{$0$}}
\put(90.00,20.00){\line(0,1){3.00}}
\put(90.00,25.00){\line(0,1){3.00}}
\put(90.00,30.00){\line(0,1){3.00}}
\put(90.00,35.00){\line(0,1){3.00}}
\put(90.00,40.00){\line(0,1){3.00}}
\put(90.00,45.00){\line(0,1){3.00}}
\put(90.00,50.00){\line(0,1){3.00}}
\put(90.00,55.00){\line(0,1){3.00}}
\put(90.00,60.00){\line(0,1){3.00}}
\put(90.00,65.00){\line(0,1){3.00}}
\put(90.00,70.00){\line(0,1){3.00}}
\put(90.00,75.00){\line(0,1){3.00}}
\put(90.00,80.00){\line(0,1){3.00}}
\put(90.00,85.00){\line(0,1){3.00}}
\put(90.00,90.00){\line(0,1){3.00}}
\put(90.00,95.00){\line(0,1){3.00}}
\put(90.00,100.00){\line(0,1){3.00}}
\put(90.00,105.00){\line(0,1){3.00}}
\put(90.00,110.00){\line(0,1){3.00}}
\put(90.00,115.00){\line(0,1){3.00}}
\put(90.00,13.00){\makebox(0,0)[cc]{${\scriptstyle  1}$}}
\put(278.89,20.00){\vector(1,0){0.2}}
\put(178.89,20.00){\line(1,0){100.00}}
\put(178.89,120.00){\vector(0,1){0.2}}
\put(178.89,20.00){\line(0,1){100.00}}
\put(172.89,115.00){\makebox(0,0)[cc]{$\gamma$}}
\put(276.89,13.00){\makebox(0,0)[cc]{$\lambda$}}
\put(172.89,13.00){\makebox(0,0)[cc]{$0$}}
\put(248.89,20.00){\line(0,1){3.00}}
\put(248.89,25.00){\line(0,1){3.00}}
\put(248.89,30.00){\line(0,1){3.00}}
\put(248.89,35.00){\line(0,1){3.00}}
\put(248.89,40.00){\line(0,1){3.00}}
\put(248.89,45.00){\line(0,1){3.00}}
\put(248.89,50.00){\line(0,1){3.00}}
\put(248.89,55.00){\line(0,1){3.00}}
\put(248.89,60.00){\line(0,1){3.00}}
\put(248.89,65.00){\line(0,1){3.00}}
\put(248.89,70.00){\line(0,1){3.00}}
\put(248.89,75.00){\line(0,1){3.00}}
\put(248.89,80.00){\line(0,1){3.00}}
\put(248.89,85.00){\line(0,1){3.00}}
\put(248.89,90.00){\line(0,1){3.00}}
\put(248.89,95.00){\line(0,1){3.00}}
\put(248.89,100.00){\line(0,1){3.00}}
\put(248.89,105.00){\line(0,1){3.00}}
\put(248.89,110.00){\line(0,1){3.00}}
\put(248.89,115.00){\line(0,1){3.00}}
\put(248.89,13.00){\makebox(0,0)[cc]{${\scriptstyle  1/D}$}}
\bezier{592}(20.00,20.00)(91.67,47.00)(88.67,118.00)
\linethickness{1pt}
\bezier{492}(20.00,20.00)(67.67,53.67)(75.33,117.67)
\bezier{236}(61.33,68.00)(80.67,76.33)(118.67,76.67)
\linethickness{0.4pt}
\bezier{276}(20.00,20.00)(38.00,60.00)(61.33,68.00)
\put(110.00,38.40){\parbox[c]{2cm}{{\small continuum}\\{\small limit}}}
\put(44.33,89.00){\makebox(0,0)[cc]{I}}
\put(97.00,49.67){\makebox(0,0)[cc]{II}}
\put(96.67,96.33){\makebox(0,0)[cc]{III}}
\linethickness{1pt}
\bezier{428}(178.89,44.33)(220.89,55.67)(233.89,117.33)
\bezier{224}(223.89,85.33)(244.89,92.00)(279.22,91.67)
\linethickness{0.4pt}
\bezier{556}(178.89,33.33)(245.22,46.33)(247.56,118.00)
\bezier{228}(178.89,53.67)(197.22,76.33)(223.89,85.33)
\put(201.56,98.00){\makebox(0,0)[cb]{I}}
\put(234.22,39.67){\makebox(0,0)[cc]{II}}
\put(255.56,104.00){\makebox(0,0)[cc]{III}}
\put(163.89,33.67){\makebox(0,0)[cc]{${\scriptstyle 2\ln(2D-1)}$}}
\put(196.59,76.00){\makebox(0,0)[cc]{{\small 1}}}
\put(233.89,123.00){\makebox(0,0)[cc]{{\small 2}}}
\put(247.56,123.00){\makebox(0,0)[cc]{{\small 3}}}
\put(282.89,91.67){\makebox(0,0)[cc]{{\small 4}}}
\put(75.33,123.00){\makebox(0,0)[cc]{{\small 2}}}
\put(88.67,123.00){\makebox(0,0)[cc]{{\small 3}}}
\put(123.33,76.67){\makebox(0,0)[cc]{{\small 4}}}
\put(34.67,52.67){\makebox(0,0)[cc]{{\small 1}}}
\put(69.00,0.00){\makebox(0,0)[cc]{a}}
\put(228.89,0.00){\makebox(0,0)[cc]{b}}
\put(29.33,21.67){\vector(-4,-1){0.2}}
\multiput(107.33,39.00)(-0.54,-0.12){145}{\line(-1,0){0.54}}
\end{picture}
\caption {
 The phase diagram of the strong coupling lattice model: a -- $D=1$, b
-- $D>1$. I -- strong coupling
(confining) phase, II, III -- weak coupling (deconfining) phases;
1 -- line on which the weak coupling phase terminates:
$\gamma=\gamma_*(\lambda)$, 2 -- line of the first--order phase
transition, 3 -- line of the instability of the strong coupling phase:
$\gamma=\gamma_c(\lambda)$, 4 -- line of the third--order large $N$ phase
transition: $\gamma=\gamma_*(\lambda)$}\label{2} \end{figure}

The consideration above gives the following picture of the phase
transition (fig.~\ref{2}). In all cases the confining and deconfining
phases can coexist and are separated by a first--order phase transition.
The model also undergoes a third--order large--$N$ phase transition in
the deconfining phase.  Actually, the line on which the weak
coupling solution terminates, $\gamma =\gamma _*(\lambda )$, obtained from
the approximate equation \rf{est}, crosses the line of the instability of
the strong coupling phase $\gamma =\gamma _c(\lambda )$ (fig.~\ref{2}).
As \rf{est} should give an upper bound for $\gamma _*(\lambda )$, this
crossing is not an artifact of the approximation done. So the weak
coupling solution terminates before the transition to the strong coupling
region and there exist two deconfining phases.
The correlation length does not turn to zero at the critical line
separating these phases. The phase transition is connected with the
large--$N$ critical behaviour of the link integral \rf{defi}. Such
 third--order phase transitions are typical for unitary matrix models
\cite{GW,JS,Wadia}.

The phase diagram of the $D=1$ model is depicted on fig. \ref{2}a. One
can verify that in the continuum limit $\gamma \rightarrow 0$, $\lambda
\rightarrow 0$ eq.  \rf{gc'} reduces to eq. \rf{gc} and eq. \rf{est}
really gives an upper bound ($\gamma _*(\lambda )\approx\frac{\pi
^4}{2}\,\lambda=48.70\, \lambda$) for \rf{g*}.

\section{Results and discussion}

 We have considered the thermodynamics of the gas of quarks in adjoint
 representation of the gauge group interacting via non-Abelian electric
 forces. In 1+1 dimension the model that we have considered is adjoint QCD
 in the limit where quarks are heavy.  The
 fugacity parameter is proportional to the Boltzmann weight of the
 classical particle of mass $m$ --
 $\lambda \propto\e^{-m/T}$. The pre--exponential
 factor is not determined by classical theory, but it
 can be computed from a loop diagram:
 \begin{equation} \lambda=\sqrt{\frac{mT}{2\pi}}\e^{-m/T}. \end{equation}
 For the classical thermodynamics to be applicable, particle mass should
 be much larger than temperature -- $m>>T$. We also assume that 
 $m^2>>e^2N$
 and pair production is suppressed. The results of Sec.~\ref{1d}  indicate
 that there exists a region of parameters in which both of the conditions
 are satisfied and the model undergoes the deconfining phase transition.
 It is of the first order with critical line given
 approximately by the equation
\begin{equation} \frac{e^2 N}{2T_c}\approx
 4.2 \sqrt{ \frac{mT_c}{2\pi} }\e^{-m/T_c}, \end{equation} or,
explicitly, \begin{equation}
T_c=\frac{m}{\ln\frac{m^2}{e^2N}+O\left(\ln\ln\frac{m^2}{e^2N}\right)}.
\end{equation} The value of $T_c$ determines, by standard arguments,
the asymptotics of the density of states in 1+1 dimensional QCD with
heavy adjoint matter -- $\rho(E)\propto\e^{E/T_c}$ for large $E$.  The
spectrum and other properties of 1+1-dimensional QCD with adjoint
matter fields have been investigated recently \cite{kut} - \cite{lst}.
As emphasized by Kutasov ~\cite{kut}, this model shares some features
with string theories in that it exhibits an infinite number of
asymptotically linear Regge trajectories and its density states
increases exponentially at high energy.  It should therefore exhibit a
Hagedorn temperature, which could either be an upper limiting
temperature, or the critical temperature of a phase transition.  Kogan
and Zhitnitsky ~\cite{kz} outlined features which the spectrum would
have to possess in order that the behavior is a phase transition.

In the multidimensional case, we consider the limit of the lattice
theory in which the coupling constant and the temperature measured in
lattice units are large. In this limit the magnetic interactions can
be neglected and the model also appears to be explicitly solvable. The
deconfining phase transition for $D>1$ takes place even in pure
gluodynamics. For the model under consideration, this corresponds to
$\lambda =0$, the case which was considered previously \cite{Zar} --
\cite{BCDMP}. The new feature of the model with adjoint matter is
appearance of the additional third--order phase transition of
Gross--Witten type for sufficiently large value of the fugacity
parameter.

\subsection*{Acknowledgments}

The work of K.Z. was supported in part by RFFI
grant 96-01-00344 and the work of G.S. was supported in part by the
Natural Sciences and Engineering Research Council of Canada.

\end{document}